\documentclass[twocolumn,showpacs,preprintnumbers,amsmath,amssymb,floatfix,prd]{revtex4}
\usepackage{epsfig,bm,lscape}
\usepackage{dcolumn}
\begin{document}
\title{On the Practical Application of Physical Anomalous Dimensions}
\author{Martin Hentschinski}\email{hentsch@bnl.gov}
\affiliation{Physics Department, Brookhaven National Laboratory, Upton, NY~11973, USA}
\author{Marco Stratmann}\email{marco@bnl.gov}
\affiliation{Physics Department, Brookhaven National Laboratory, Upton, NY~11973, USA}

\begin{abstract}
We revive the idea of using physical anomalous dimensions in the QCD scale evolution of deep-inelastic structure functions 
and their scaling violations and present a detailed phenomenological study of its applicability.
Differences with results obtained in the conventional framework of scale-dependent
quark and gluon densities are discussed and traced back to the
truncation of the perturbative series at a given order in the strong coupling. 
\end{abstract}


\maketitle

%%%%%%%%%%%%%%%%%%%%%%%%%%%%%%%%%%%%%
\section{Introduction and Motivation}
%%%%%%%%%%%%%%%%%%%%%%%%%%%%%%%%%%%%%
%
Measurements of deep-inelastic scattering (DIS) cross sections 
are routinely analyzed in terms of scale-dependent quark and gluon distribution functions $f_i(x,Q^2)$,
$i=q,\bar{q},g$ in global QCD fits \cite{ref:pdffits}.
In particular, the very accurate DIS data from the DESY-HERA experiments \cite{Aaron:2009aa} 
provide the backbone of such type of QCD analyses, also thanks to their vast coverage in the relevant 
kinematic variables $x$ and $Q^2$, denoting the momentum fraction carried by the struck parton 
and the resolution scale set by the momentum transfer squared, respectively.

The conventional theoretical framework for DIS is based on the factorization theorem \cite{Collins:1989gx},
which allows one to organize the computation of DIS structure functions $F_{2,L}(x,Q^2)$ as a convolution of 
perturbatively calculable Wilson coefficients \cite{Furmanski:1981cw,vanNeerven:1991nn,Moch:1999eb,Kazakov:1987jk} 
and parton distribution functions (PDFs) capturing the long-distance, 
non-perturbative physics \cite{Blumlein:2012bf}. 
Similar ideas are successfully applied to more complicated hard-scattering 
processes studied at hadron-hadron colliders, 
which provide further, invaluable constraints on PDFs in global QCD analyses \cite{ref:pdffits}.

To make the separation between long- and short-distance physics manifest, one needs to 
introduce some arbitrary factorization scale $\mu_f$, apart from the scale $\mu_r$ 
appearing in the renormalization of the strong coupling $\alpha_s$.
The independence of physical observables such as $F_{2,L}$ on $\mu_f$ can be used to derive powerful
renormalization group equations (RGEs) governing the scale dependence of PDFs in each order
of perturbation theory. The corresponding kernels are the anomalous dimensions or 
splitting functions associated with collinear two-parton configurations \cite{ref:lo,ref:nlo,ref:nnlo}.
Since factorization can be carried out in infinitely many different ways, one is left with an additional
choice of the factorization scheme for which one usually adopts the $\overline{\mathrm{MS}}$
prescription \cite{Bardeen:1978yd}. Likewise, the RGE governing the running of $\alpha_s$ with $\mu_r$ can be deduced from 
taking the derivative of $F_{2,L}$ with respect to $\mu_r$.  
Upon properly combining PDFs and Wilson coefficients in the same factorization scheme, 
any residual dependence on $\mu_f$ is suppressed by an additional power of $\alpha_s$,
i.e., is formally one order higher in the perturbative expansion but not necessarily numerically small.

Alternatively, it is possible to formulate QCD scale evolution equations directly for physical
observables without resorting to auxiliary, convention-dependent quantities such as PDFs.
This circumvents the introduction of a factorization scheme and $\mu_f$ and, hence, any dependence of the results
on their actual choice. 
The concept of physical anomalous dimensions is not at all a new idea 
and has been proposed quite some time ago \cite{Furmanski:1981cw,Catani:1996sc,Blumlein:2000wh}
but its practical aspects have never been studied in detail.
The framework is suited best for theoretical analyses based on DIS data 
with the scale $\mu_r$ in the strong coupling being the only theoretical ambiguity.
In addition, $F_{2,L}$ or their scaling violations can be parametrized much
more economically than a full set of quark and gluon PDFs, which greatly simplifies any 
fitting procedure and phenomenological analysis. 
The determination of $\alpha_s$ from fits to DIS structure functions
is the most obvious application, as theoretical scheme and scale uncertainties are reduced to a minimum.
We note that physical anomalous dimension were used, however, also as a calculational 
tool to study, e.g., all-order aspects of the perturbative series 
for splitting and coefficient functions in the limit of large momentum fractions $x$ \cite{Soar:2009yh}.

Recently, the interest in DIS has been revived in a series of dedicated studies in search of a 
compelling physics case for a 
future high-luminosity electron-ion collider, such as the proposed
EIC \cite{Boer:2011fh} and LHeC \cite{AbelleiraFernandez:2012ni} projects.
One of the driving physics goals is a detailed mapping of the transition 
into a non-linear kinematic regime dominated by high, or saturated, gluon densities
by measuring, for instance, $F_{2,L}$ and their scaling violations very precisely at small $x$ 
both in electron-proton and in electron-heavy ion collisions. 
A signature for the onset of saturation effects would be the observation of
deviations from the linear, Dokshitzer-Gribov-Lipatov-Altarelli-Parisi (DGLAP) type
of QCD scale evolution, which, again, could be identified best by performing analyses 
in the framework of physical anomalous dimensions.

In this paper we will largely focus on the practical implementation of physical anomalous
dimensions in analyses of DIS data up to next-to-leading order (NLO) accuracy. 
We shall study in detail potential differences with results 
obtained in the conventional framework based on scale-dependent quark and gluon densities,
which could be caused by the way how the perturbative series is truncated at any given order.
To the best of our knowledge, these practical aspects have never been studied before, apart
from a brief application of physical anomalous dimensions in an attempt to extract the
strong coupling from DIS structure function data taken with longitudinally polarized leptons and hadrons 
\cite{Bluemlein:2002be}. 

The results obtained in this note should provide the framework for 
using physical anomalous dimensions in actual quantitative 
phenomenological studies to be pursued once data from a future electron-ion collider
become available or for projections based on pseudo-data.
Apart from the applications related to gluon saturation, 
precise data for the longitudinally polarized DIS structure function
$g_1$ and its scaling violations $dg_1/d\ln Q^2$  
are expected from an EIC \cite{Boer:2011fh,Aschenauer:2012ve}.
Such data should allow one to revisit the attempted extraction of $\alpha_s$ 
performed in Ref.~\cite{Bluemlein:2002be}.
We note that physical anomalous dimensions can be also applied to time-like processes
such as inclusive hadron production in electron-positron annihilation. These processes
are traditionally parametrized by fragmentation functions, which account for the
non-perturbative transition of a certain quark flavor or gluon into the observed 
hadron species \cite{deFlorian:2007aj}.
Again it is in principle possible to use time-like equivalents of DIS structure functions
and evolve them in a factorization scheme independent way. 

The paper is organized as follows: in Section II we briefly recall how to derive physical
anomalous dimensions to establish our notation and conventions used throughout the paper.
Explicit results are given in leading and next-to-leading order of perturbation theory
and compared with the results known in the literature.
In Section III we discuss the numerical implementation, outline subtleties related
to the truncation of the perturbative series at NLO, and show how to circumvent them.
We briefly summarize our results in Section IV.

%%%%%%%%%%%%%%%%%%%%%%%%%%%%%%%%%%%%%%%%%%%%%%%%%%%%%%
\section{Theoretical Framework \label{sec:theory}}
%%%%%%%%%%%%%%%%%%%%%%%%%%%%%%%%%%%%%%%%%%%%%%%%%%%%%%
%
This gist of the factorization scheme-invariant framework
amounts to combine any two DIS observables $\{F_A,F_B\}$ 
and determine their corresponding $2\times 2$ matrix of physical anomalous dimensions
in lieu of the scale-dependent quark singlet, $\Sigma\equiv \sum_q (q+\bar{q})$,
and gluon distributions appearing in the standard, coupled singlet DGLAP evolution equations.
Instead of using measurements of $F_2$ and $F_L$ (actually their flavor singlet parts), 
one can also utilize their variation with 
scale for any given value of $x$, i.e., $dF_{2,L}(x,Q^2)/d\ln Q^2$ as an observable. 
The required sets of physical anomalous dimensions for both $\{F_2,F_L\}$ and
$\{F_2,dF_2/d\ln Q^2\}$ have been derived in \cite{Blumlein:2000wh} up to NLO
accuracy.
The additionally needed evolution equations for the non-singlet portions of
the structure functions $F_{2,L}$ are simpler and not matrix valued. 

%%%%%%%%%%%%%%%%%%%%%%%%%%%%%%%%%%%%%%%%%%%%%%%%%%%%%%%
\subsection{Physical Anomalous Dimensions \label{sec:physanom}}
%%%%%%%%%%%%%%%%%%%%%%%%%%%%%%%%%%%%%%%%%%%%%%%%%%%%%%%
%
As we shall see below, the required physical anomalous dimensions comprise the inverse of
coefficient and splitting functions and are most conveniently expressed
in Mellin $n$ moment space.
The Mellin transformation of a function $\phi$ given in
Bjorken $x$ space, such as PDFs or splitting functions, is defined as
\begin{equation}
\label{eq:mellin}
\phi(n) \equiv \int_0^1 dx \,x^{n-1} \phi(x)\;,
\end{equation}
where $n$ is complex valued. 
As an added benefit, convolutions in $x$ space turn into ordinary products upon
applying (\ref{eq:mellin}), which, in turn, allows for an analytic solution of QCD scale
evolution equations for PDFs.
The corresponding inverse Mellin transformation
is straightforwardly performed numerically along a suitable 
contour in $n$ space, see, e.g., Ref.~\cite{Vogt:2004ns} for details.
The necessary analytic continuations to non-integer $n$ moments are given in 
\cite{Gluck:1989ze,Blumlein:2000hw}, and an extensive 
list of Mellin transforms is tabulated in \cite{Blumlein:1998if}.
We will work in Mellin space throughout this paper.

Assuming factorization, moments of DIS structure functions $F_I$ at a scale $Q$
can be expressed as
\begin{align}
\label{eq:strfct}
F_{I}(n,Q^2)  = 
\sum_{k=q,\bar{q},g} &  e_k^2\,  C_{I,k}    \left (n,a_s(\mu_r^2),\frac{Q^2}{\mu_f^2}, \frac{\mu_r^2}{\mu_f^2} \right)     
\, \notag \\ 
%&  \qquad  \qquad \cdot    \,  
&\cdot
f_k\left(n,a_s(\mu_r^2),\frac{\mu_f^2}{Q_0^2}, \frac{\mu_r^2}{\mu_f^2} \right) 
\end{align}
where the sum runs over all contributing $n_f$ active quark flavors with electric charge
$e_q$ and the gluon $g$, each represented by a PDF $f_k$.  
For $e_g^2$, the averaged quark charge factor $\bar{e}_f^2 = (1/n_f) \sum_q e_q^2$ has to be used instead.  
$\mu_r$ and $\mu_f$ specify renormalization and
factorization scale, respectively. The scale $Q_0$ defines the starting scale for
the PDF evolution, where a set of non-perturbative input distributions
needs to be specified.  For simplicity we identify in the following
the renormalization scale with the factorization scale, i.e., $\mu_r = \mu_f
\equiv \mu$.  The coefficient functions $C_{I,k}$ are calculable in pQCD
\cite{Furmanski:1981cw,vanNeerven:1991nn,Moch:1999eb,Kazakov:1987jk}
and exhibit the following series in $a_s\equiv \alpha_s/4\pi$
\begin{equation}
\label{eq:cexpansion}
C_{I,k}\left(n,a_s(\mu^2),\frac{Q^2}{\mu^2}  \right) =  \sum_{m = m_0} a^{m}_s(\mu^2) \; C^{(m)}_{I,k}\left(n, \frac{Q^2}{\mu^2} \right) \;,
\end{equation}
where $m_0$ depends on the first non-vanishing order in $a_s$ in the expansion for the observable
under consideration, e.g.,
$m_0=0$ for $F_2$ and $m_0=1$ for $F_L$. 

Each PDF $f_k(n,\mu^2/Q_0^2)$ obeys the DGLAP evolution equation which reads 
\begin{equation}
\label{eq:dglap}
\frac{d}{d\ln \mu^2}  f_k\left(n,\frac{\mu^2}{Q_0^2} \right)  = \sum_{l=q,\bar{q},g} P_{kl} \left (n,a_s(\mu^2) \right) f_l\left(n, \frac{\mu^2}{Q_0^2} \right)
\end{equation} 
where the $l\to k$ splitting functions have a similar expansion \cite{ref:lo,ref:nlo,ref:nnlo} as the coefficient
functions in Eq.~(\ref{eq:cexpansion}):
\begin{equation}
\label{eq:pexpansion}
P_{kl}(n,a_s(\mu^2))  =  \sum_{m = 0}  a_s(\mu^2)^{1 + m} \; P^{(m)}_{kl}(n) \;.
\end{equation}
The $P_{kl}(n)$ 
relate to the corresponding anomalous dimensions through $\gamma_{kl}(n)=-2P_{kl}(n)$
in the normalization conventions we adopt, where we use the leading order (LO) and NLO expressions for $\gamma_{kl}(n)$
given in App.~B of the first reference in \cite{ref:nlo}.
We note that the same normalization is used in the publicly available {\sc Pegasus} evolution code \cite{Vogt:2004ns}.
In practice one distinguishes a $2\times 2$ matrix-valued DGLAP equation evolving the flavor singlet 
vector comprising $\Sigma(n,\mu^2/Q_0^2)$ and $g(n,\mu^2/Q_0^2)$
%, with
%
%\begin{align}
%  \label{eq:singlet}
%  \Sigma(n, \mu^2) & = \sum_{f = 1}^{n_f} \left(  q_f(n, \mu^2) + \bar{q}_f(n, \mu^2) \right)
%\end{align}
%
and a set of $n_f-1$ RGEs for the relevant non-singlet quark flavor combinations. 

The scale-dependent strong coupling itself obeys another RGE governed by the QCD beta function
\begin{equation}
\label{eq:alphas}
\frac{da_s(\mu)}{d\ln \mu^2} = \beta(a_s) = - \sum_m a_s^{m+2} \beta_m
\end{equation}
with $\beta_0=11-2n_f/3$ and $\beta_1=102-38 n_f/3$ up to NLO accuracy.
To compare below with the results for the physical anomalous dimensions in Ref.~\cite{Blumlein:2000wh}
we also introduce the evolution variable
\begin{equation}
\label{eq:t}
t \equiv - \frac{2}{\beta_0} \ln \left( \frac{a_s (Q^2)}{a_s (Q_0^2)} \right)\;.
\end{equation}

Instead of studying $F_I(n,Q^2)$ in (\ref{eq:strfct})
in terms of scale-dependent PDFs, which are obtained from solving 
the singlet and non-singlet combinations of DGLAP equations (\ref{eq:dglap})
in a fit to data \cite{ref:pdffits}, one can also derive
evolution equations directly in terms of the observables $F_I(n,Q^2)$.
To this end, we consider a pair of DIS observables $F_A$ and $F_B$, 
to be specified below,
whose scale dependence is governed by a coupled matrix-valued equation 
\begin{align}
\label{eq:physsing}
& \frac{d}{d\ln Q^2} 
\begin{pmatrix}
      F_A^{(S)}(n, Q^2) \\ 
      F_B^{(S)}(n, Q^2)
\end{pmatrix}= \notag \\
&\begin{pmatrix}
      K_{AA} & K_{AB} \\ 
      K_{BA} & K_{BB}
	\end{pmatrix}  (n,a_s(Q^2))  \cdot 
\begin{pmatrix}
      F_A^{(S)} (n, Q^2) \\ 
      F_B^{(S)} (n, Q^2)
\end{pmatrix}	
\end{align} 
for the flavor singlet (S) parts of $F_{A,B}$
and a set of non-singlet (NS) equations
\begin{equation}
\label{eq:physnonsing}
\frac{dF_{A,B}^{(NS)}(n,Q^2)}{d\ln Q^2} = K^{(NS)}(n,a_s(Q^2)) \cdot F_{A,B}^{(NS)}(n, Q^2)
\end{equation}
for the remainders.

The required physical anomalous dimensions in Eqs.~(\ref{eq:physsing}) and (\ref{eq:physnonsing}),
obey a similar perturbative expansion in $a_s$ as in (\ref{eq:pexpansion}).
%
%%\begin{equation}
%%\label{eq:kexpansion}
%%K_{ij}(n,\alpha_s(\mu^2))  =  \sum_{m = 0}  a_s(\mu^2)^{1 + m} \; K^{(m)}_{ij}(n) \; ,
%%\end{equation}
%
The singlet kernels in (\ref{eq:physsing}) are constructed by
substituting
\begin{align}
  \label{eq:matrixproduct}
  \begin{pmatrix}
      F_A^{(S)}  (n, Q^2) \\ 
      F_B^{(S)}  (n, Q^2)
\end{pmatrix}
&= \bar{e}_f^2\,
C \left(n,a_s(\mu^2), \frac{Q^2}{\mu^2}\right) \cdot 
\begin{pmatrix}
  \Sigma   (n, {\mu^2/Q_0^2}) \\ g  (n, {\mu^2/Q_0^2})
\end{pmatrix}
\end{align}
into the left-hand side of 
Eq.~(\ref{eq:physsing}) and taking the derivatives.
Note that we have normalized the quark singlet part of $F_{A,B}$ with the same
averaged charge factor $\bar{e}_f^2$ which appears in the gluonic sector.
Upon making use of the RGEs for PDFs and the strong coupling in Eq.~(\ref{eq:dglap})
and (\ref{eq:alphas}), respectively, one arrives at
\begin{widetext}
  \begin{align}
\label{eq:physanom}
%\nonumber
% \lefteqn{K_{ij}(n,\alpha_s(Q^2)) =   }      \\
% & &  \bigg[ \bigg(
% \beta(a_s (Q^2))   \frac{ \partial C (n, \alpha_s(Q^2), 1)}{\partial  a_s (Q^2)} + C  (n, \alpha_s(Q^2), 1)  \notag \\
% &&   \hspace{1cm} \cdot  P (n, a_s(Q^2)) \bigg )   \cdot C^{-1} (n, \alpha_s(Q^2), 1)
% \bigg]_{ij}
 K_{ij}  (n,a_s(Q^2))  =   \bigg[ \bigg(
\beta(a_s (Q^2))   \frac{ \partial C (n, a_s(Q^2), 1)}{\partial  a_s (Q^2)}   
   + C  (n, a_s(Q^2), 1)  \cdot  P (n, a_s(Q^2)) \bigg )   \cdot C^{-1} (n, a_s(Q^2), 1)\,,
 \bigg]_{ij} % C^{-1} (n, \alpha_s(Q^2), 1)    
\end{align}
\end{widetext}
where we have introduced $2\times 2$ matrices 
\begin{align}
  \label{eq:coeffmatrix}
 {  C}  & =
  \begin{pmatrix}
    C_{A,q} &    C_{A,g} \\
  C_{B,q} &    C_{B,g} 
  \end{pmatrix},
&
P & =
\begin{pmatrix}
  P_{qq} & 2 n_f  P_{qg} \\
P_{gq} & P_{gg}
\end{pmatrix}.
\end{align}
for the relevant singlet coefficient and splitting functions, respectively.
An analogous, albeit much simpler expression holds for the NS kernel $K^{(NS)}$ in (\ref{eq:physnonsing}).
As has been demonstrated in \cite{Blumlein:2000wh}, the kernels (\ref{eq:physanom}) are independent of the chosen 
factorization scheme and scale but do depend on $\mu_r$ and the details of the renormalization procedure.
We also note that the inverse $C^{-1}$ in (\ref{eq:physanom}), appearing upon re-expressing all PDFs by $F_{A,B}$,
can be straightforwardly computed only in Mellin moment space.

%%%%%%%%%%%%%%%%%%%%%%%%%%%%%%%%%%%%%%%
\subsection{Example I: $F_2$ and $F_L$ \label{sec:exp1}}
%%%%%%%%%%%%%%%%%%%%%%%%%%%%%%%%%%%%%%%
%
Let us first consider the evolution of the 
pair of observables $\{F_2,F_L\}$, both of which can be
obtained from measurements of the reduced DIS cross section
at different energies through a Rosenbluth separation.
A precise determination of $F_L$ in a broad kinematic regime 
is a key objective at both 
an EIC \cite{Boer:2011fh} and the LHeC \cite{AbelleiraFernandez:2012ni}.

Since the perturbative series for $F_L$ only starts at ${\cal{O}}(a_s)$,
one wants to account for this offset by actually considering the evolution of 
either $\{F_2,F_L/(a_s C_{L,g}^{(1)}) \}$ or $\{F_2,F_L/(a_s C_{L,q}^{(1)}) \}$.
Both sets of kernels $K_{AB}$ show a rather different behavior with $n$, as we
shall illustrate below, but without having any impact 
on the convergence properties of the inverse Mellin transform
needed to recover the $x$ dependent structure functions.
The kernels $K_{AB}$ at LO and NLO accuracy for $\{F_A,F_B\}=\{F_2,F_L/(a_s C_{L,g}^{(1)}) \}$
can be found in \cite{Blumlein:2000wh}.
Note that evolution in \cite{Blumlein:2000wh} is expressed in terms of $t$.
Using (\ref{eq:t}), $d/da_s=-2/(a_s \beta_0) d/dt$, and (\ref{eq:alphas}) 
to compute the extra terms proportional to $\beta(a_s)$,
we fully agree with their results.

For $\{F_A,F_B\}=\{F_2,F_L/(a_s C_{L,q}^{(1)}) \}$ one finds
\begin{align}
\nonumber
K_{22}^{(0)} &= P_{qq}^{(0)}-\frac{C_{L,q}^{(1)} }{C_{L,g}^{(1)}} P_{qg}^{(0)} ,\;\;\;
K_{2L}^{(0)} =\frac{C_{L,q}^{(1)} }{C_{L,g}^{(1)}} P_{qg}^{(0)} ,\\
\nonumber
K_{L2}^{(0)} &=  \left(\frac{C_{L,g}^{(1)} }{C_{L,q}^{(1)}}-\frac{C_{L,q}^{(1)}
  }{C_{L,g}^{(1)}} \right)  P_{qg}^{(0)}    -P_{gg}^{(0)}+P_{qq}^{(0)},\\
K_{LL}^{(0)} &= \frac{C_{L,q}^{(1)} }{C_{L,g}^{(1)}}  P_{qg}^{(0)}  + P_{gg}^{(0)}
\label{eq:klo-clq}
\end{align}
at the LO approximation, i.e., after expanding Eq.~(\ref{eq:physanom}) up to ${\cal{O}}(a_s)$.
Only the off-diagonal entries change if dividing $F_L$ by $a_s C_{L,g}^{(1)}$; see 
Eqs.~(41)-(45) in Ref.~\cite{Blumlein:2000wh}.
Likewise, at NLO accuracy we obtain
\begin{align}
\nonumber
K_{22}^{(1)} & = 
P_{{qq}}^{(1)}
 -\frac{C_{{L,q}}^{(1)} }{C_{{L,g}}^{(1)}}P_{{qg}}^{(1)}
 + \frac{C_{{2,g}}^{(1)} C_{{L,q}}^{(1)} }{C_{{L,g}}^{(1)}}  P_{{qq}}^{(0)}
+C_{{2,g}}^{(1)} P_{{gq}}^{(0)}
 \notag \\
& 
 + \left[ \frac{C_{{L,g}}^{(2)} C_{{L,q}}^{(1)}}{ \left( C_{{L,g}}^{(1)} \right)^2}
 - C_{{2,g}}^{(1)} \left( \frac{ C_{{L,q}}^{(1)}}{C_{{L,g}}^{(1)}} \right)^2
 - \frac{C_{{L,q}}^{(2)}}{C_{{L,g}}^{(1)}}
\right]
 P_{{qg}}^{(0)}
\notag \\
&
-\frac{C_{{2,g}}^{(1)} C_{{L,q}}^{(1)} P_{{gg}}^{(0)}}{C_{{L,g}}^{(1)}}
                + \beta _0 \bigg( \frac{ C_{{2,g}}^{(1)} C_{{L,q}}^{(1)}}{C_{{L,g}}^{(1)}}  -  C_{{2,q}}^{(1)}\bigg)
\end{align}
\begin{align}
\nonumber
K^{(1)}_{2L} &=   
                    \frac{C_{{L,q}}^{(1)} }{C_{{L,g}}^{(1)}}P_{{qg}}^{(1)}
                    - \frac{C_{{2,g}}^{(1)} C_{{L,q}}^{(1)} }{C_{{L,g}}^{(1)}}P_{{qq}}^{(0)}
                    + \frac{C_{{2,g}}^{(1)} C_{{L,q}}^{(1)} }{C_{{L,g}}^{(1)}}P_{{gg}}^{(0)} \\
&
\nonumber
                     + \left[ C_{{2,g}}^{(1)} \left( \frac{ C_{{L,q}}^{(1)}}{C_{{L,g}}^{(1)}} \right)^2
                     + \frac{C_{{2,q}}^{(1)} C_{{L,q}}^{(1)}}{C_{{L,g}}^{(1)}}
                     - \frac{C_{{L,g}}^{(2)} C_{{L,q}}^{(1)}}{\left(C_{{L,g}}^{(1)}\right)^2} \right]  P_{{qg}}^{(0)} \\
&
\nonumber
                  - \beta_0 \frac{ C_{{2,g}}^{(1)} C_{{L,q}}^{(1)}}{C_{{L,g}}^{(1)}}
                 \,,
\end{align}
\begin{align}
\nonumber
K_{L2}^{{(1)}} & =     \left(\frac{C_{{L,g}}^{(1)}}{C_{{L,q}}^{(1)}}
                      -\frac{C_{{L,q}}^{(1)}}{C_{{L,g}}^{(1)}} \right) P_{{gq}}^{(1)}
                     -P_{{gg}}^{(1)} +P_{{qq}}^{(1)} \\
\nonumber
&
                      +\left( \frac{C_{{2,g}}^{(1)} C_{{L,q}}^{(1)} }{C_{{L,g}}^{(1)}}
                       -C_{{2,q}}^{(1)} 
                       +\frac{C_{{L,q}}^{(2)}}{C_{{L,q}}^{(1)}}  \right)   P_{{qq}}^{(0)}
 \\
\nonumber
&
                + \bigg( 
                         \frac{C_{{L,g}}^{(2)}}{C_{{L,q}}^{(1)}}                                - \frac{C_{{2,q}}^{(1)} C_{{L,g}}^{(1)} }{C_{{L,q}}^{(1)}} 
                        +  C_{{2,g}}^{(1)}
                    \bigg) P_{{gq}}^{(0)} \\
\nonumber
& 
                    + \left( 
                          C_{{2,q}}^{(1)}  
                         -\frac{C_{{2,g}}^{(1)} C_{{L,q}}^{(1)} }{C_{{L,g}}^{(1)}}
                      - \frac{C_{{L,q}}^{(2)} }{C_{{L,q}}^{(1)}}
                      \right)  P_{{gg}}^{(0)} \\
\nonumber
&                  
               + \bigg[ 
                      \frac{C_{{2,q}}^{(1)} C_{{L,q}}^{(1)} }{C_{{L,g}}^{(1)} } 
                      -  C_{{2,g}}^{(1)} \left( \frac{ C_{{L,q}}^{(1)}}{C_{{L,g}}^{(1)}} \right)^2                    
                      -\frac{2 C_{{L,q}}^{(2)} }{C_{{L,g}}^{(1)}}                 
 \\
\nonumber
&
                   + \frac{C_{{L,g}}^{(2)} C_{{L,q}}^{(1)} }{ \left(C_{{L,g}}^{(1)} \right)^2}
                  \bigg] P_{{qg}}^{(0)}     
                   + \beta_0 \left( \frac{ C_{{L,g}}^{(2)}}{C_{{L,g}}^{(1)}} 
                      -\frac{ C_{{L,q}}^{(2)}}{C_{{L,q}}^{(1)}} \right )
                     \,,
\end{align}
\begin{align}
\nonumber
K_{LL}^{{(1)}} & =  \frac{C_{{L,q}}^{(1)}}{C_{{L,g}}^{(1)}} P_{{qg}}^{(1)}
                 +P_{{gg}}^{(1)}
                  -\frac{C_{{2,g}}^{(1)} C_{{L,q}}^{(1)} }{C_{{L,g}}^{(1)}}P_{{qq}}^{(0)} \\
\nonumber 
&
            + \bigg[ 
              C_{{2,g}}^{(1)} \left( \frac{ C_{{L,q}}^{(1)} }{C_{{L,g}}^{(1)}}\right)^2
                -\frac{C_{{L,g}}^{(2)} C_{{L,q}}^{(1)} }{ \left(C_{{L,g}}^{(1)}\right)^2}
                 +\frac{C_{{L,q}}^{(2)} }{C_{{L,g}}^{(1)}}
               \bigg] P_{{qg}}^{(0)} \\
\label{eq:knlo-clq}
&
+  \frac{C_{{2,g}}^{(1)} C_{{L,q}}^{(1)} }{C_{{L,g}}^{(1)}}P_{{gg}}^{(0)}
                     -C_{{2,g}}^{(1)} P_{{gq}}^{(0)}
 - \beta _0\frac{ C_{{L,g}}^{(2)}}{C_{{Lg}}^{(1)}}  \,.
\end{align}
The corresponding NS kernels read
\begin{align}
\nonumber
K_{NS}^{(0)} & = P_{qq}^{(0)}\,, \\
K_{NS}^{(1)} & = - \beta_0 C_{2,q}^{(1)}  + P_{NS}^{(1)}\,.
\end{align}

%%%%%%%%%%%%%%%%%%%%%%%%%%%%%%%%%%%%%%%
% FIGURE 1: KERNELS
%%%%%%%%%%%%%%%%%%%%%%%%%%%%%%%%%%%%%%%
%
\begin{figure}[h]
\begin{center}
\vspace*{-0.6cm}
\epsfig{figure=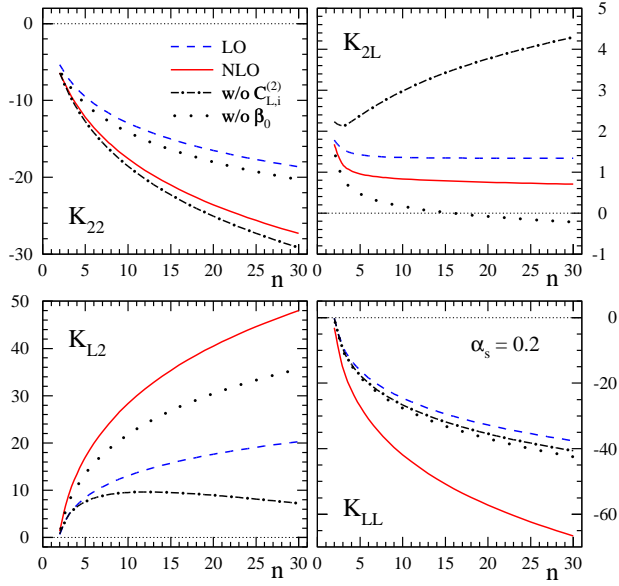,width=0.50\textwidth}
\end{center}
\vspace*{-0.5cm}
\caption{\label{fig:klo-knlo} [color online] Physical anomalous dimensions $K_{AB}(n)$ in
LO and NLO for $\{F_A,F_B\}=\{F_2,F_L/(a_s C_{L,q}^{(1)}) \}$ assuming $\alpha_s=0.2$ and $n_f=3$.
The dash-dotted and dotted lines, respectively, show the NLO results where all contributions from $C_{L,i}^{(2)}$ and $\beta_0$
to $K_{AB}^{(1)}$ in Eq.~(\ref{eq:knlo-clq}) have been omitted.
A global factor of $a_s$ has been ignored in the perturbative expansion,
i.e., $K_{AB}=K_{AB}^{(0)}+ a_s K_{AB}^{(1)}+{\cal{O}}(a_s^2)$ is displayed.}
\end{figure}
In Fig.~\ref{fig:klo-knlo} we illustrate the $n$ dependence of the LO and NLO singlet kernels $K_{AB}$
for the evolution of $\{F_A,F_B\}=\{F_2,F_L/(a_s C_{L,q}^{(1)}) \}$ assuming $\alpha_s=0.2$ and $n_f=3$. 
As can be seen, NLO corrections are sizable for all singlet kernels, in particular, when compared
with the perturbative expansion of the singlet splitting functions $P_{kl}(n)$ in (\ref{eq:pexpansion}); 
see Figs.~1 and 2 in Ref.~\cite{ref:nnlo}.
This is, however, not too surprising given that the known large higher order QCD corrections to the
Wilson coefficients $C_{L,g}$ and $C_{L,q}$ \cite{Moch:2004xu} 
are absorbed into the physical anomalous dimensions $K_{AB}$ for the evolution of 
the DIS structure functions $F_2$ and $F_L$.
The impact of contributions from the NLO coefficients $C_{L,g}^{(2)}$ and $C_{L,q}^{(2)}$
on the results obtained for $K_{AB}$
is illustrated by the dash-dotted lines in Fig.~\ref{fig:klo-knlo}.
Another source for large corrections are the terms proportional to $\beta_0$ in Eq.~(\ref{eq:knlo-clq})
as can be inferred from the dotted lines; note that $K_{L2}^{(1)}$ and $K_{LL}^{(1)}$ include terms
proportional to $\beta_0 C_{L,g}^{(2)}$ and $\beta_0 C_{L,q}^{(2)}$.
In Sec.~\ref{sec:pheno} we will demonstrate how the differences between the LO and NLO kernels
become apparent in the scale evolution of $F_{2,L}(x,Q^2)$.

%%%%%%%%%%%%%%%%%%%%%%%%%%%%%%%%%%%%%%%
% FIGURE 2: COMPARE OFF-DIAG KERNELS
%%%%%%%%%%%%%%%%%%%%%%%%%%%%%%%%%%%%%%%
%_
\begin{figure}[t!]
\begin{center}
\vspace*{-0.6cm}
\epsfig{figure=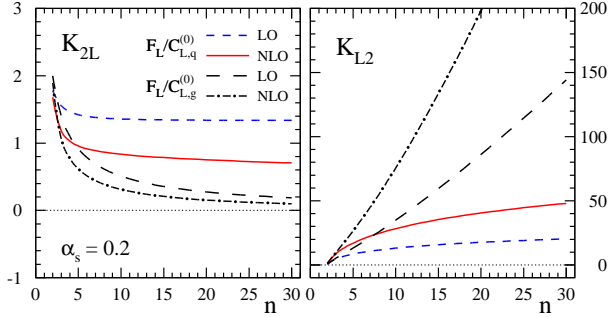,width=0.50\textwidth}
\end{center}
\vspace*{-0.5cm}
\caption{\label{fig:k-comp} [color online] As in Fig.~\ref{fig:klo-knlo} but now comparing the
LO and NLO off-diagonal kernels $K_{2L}$ and $K_{L2}$ for $\{F_A,F_B\}=$
$\{F_2,F_L/(a_s C_{L,q}^{(1)}) \}$ and $\{F_2,F_L/(a_s C_{L,g}^{(1)}) \}$; see text.}
\end{figure}
Figure~\ref{fig:k-comp} compares the LO and NLO off-diagonal kernels $K_{2L}$ and $K_{L2}$ for $\{F_A,F_B\}=$
$\{F_2,F_L/(a_s C_{L,q}^{(1)}) \}$ and $\{F_2,F_L/(a_s C_{L,g}^{(1)}) \}$. 
The most noticeable difference is the strong rise with $n$ for the kernel $K_{L2}$ governing the 
evolution of $\{F_2,F_L/(a_s C_{L,g}^{(1)}) \}$.
At LO accuracy, this is readily understood by inspecting the $n\to \infty$ limit which yields, see Eq.~(41) in
\cite{Blumlein:2000wh}, $K_{L2}^{(0)}\sim n \ln n$, recalling that asymptotically
$C_{L,q}^{(1)}\sim 1/n$, $C_{L,g}^{(1)}\sim 1/n^2$,
$P_{qq}^{(0)}\sim \ln n$, $P_{qg}^{(0)}\sim 1/n$, $P_{gq}^{(0)}\sim 1/n$, and $P_{gq}^{(0)}\sim \ln n$.
The NLO kernel $K_{L2}$ exhibits an even stronger rise with $n$.
In the same way one obtains, for instance, that $K_{L2}^{(0)}$ governing the 
evolution of $\{F_2,F_L/(a_s C_{L,q}^{(1)}) \}$ only grows like $\ln n$, see Eq.~(\ref{eq:klo-clq}).

Despite this peculiar $n$ dependence and the differences between the singlet kernels 
shown in Fig.~\ref{fig:k-comp}, both sets of observables, 
$\{F_2,F_L/(a_s C_{L,q}^{(1)}) \}$ and $\{F_2,F_L/(a_s C_{L,g}^{(1)}) \}$
can be used interchangeably in an analysis at LO and NLO accuracy.
Results for the QCD scale evolution are identical, and
one does not encounter any numerical instabilities related to the inverse Mellin transform,
which we perform along a contour as described in Ref.~\cite{Vogt:2004ns}.
In fact, it is easy to see that the eigenvalues 
\begin{equation}
\label{eq:eigenvalue}
\lambda_{\pm} = \frac{1}{2} \left[ K_{22}^{(0)} + K_{LL}^{(0)} \pm
\sqrt{\!(K_{22}^{(0)} - K_{LL}^{(0)})^2 \! + 4 K_{2L}^{(0)} K_{L2}^{(0)}}\,\right],
\end{equation}
which appear when solving the matrix valued evolution equation (\ref{eq:physsing}),
are identical for both sets of kernels and also agree with the corresponding
eigenvalues for the matrix of singlet anomalous dimensions $P^{(0)}_{kl}$; see also the discussions
in Sec.~\ref{sec:pheno}

%%%%%%%%%%%%%%%%%%%%%%%%%%%%%%%%%%%%%%%
\subsection{Example II: $F_2$ and $dF_2/dt$ \label{sec:exp2}}
%%%%%%%%%%%%%%%%%%%%%%%%%%%%%%%%%%%%%%%
%
Of future phenomenological interest could be also the
pair of observables $\{F_2,dF_2/dt\}$, in particular, in the 
absence of precise data for $F_L$.
Determining experimentally the $t$ or $Q^2$ slope of $F_2$ is,
of course, also challenging.

Defining $F_D\equiv dF_2/dt$, we obtain the following
physical evolution kernels
\begin{align}
\label{eq:klo-deriv}
K_{22}^{(0)} & = 0\,, \hspace{0.5cm} 
K_{2D}^{(0)} = 2\,, \nonumber \\
K_{D2}^{(0)} & = \frac{1}{2} \left[P_{{gq}}^{(0)} P_{{qg}}^{(0)}
               - P_{{gg}}^{(0)} P_{{qq}}^{(0)}\right] \,, 
\nonumber \\ 
K_{2D}^{^{(0)}} & = P_{\text{gg}}^{(0)}+P_{\text{qq}}^{(0)}
\end{align}  
at LO and
\begin{align}
\label{eq:knlo-deriv}
K_{22}^{(1)} & = 0\,, \hspace{0.5cm}
K_{2D}^{(1)} = \frac{2 \beta _1}{\beta _0}\,,  
\nonumber \\
  K_{D2}^{(1)} & = \frac{1}{2}  \bigg\{
                    P_{{gq}}^{(1)} P_{{qg}}^{(0)}   +  P_{{gq}}^{(0)} P_{{qg}}^{(1)}
                    - P_{{gg}}^{(1)} P_{{qq}}^{(0)}   -  P_{{gg}}^{(0)} P_{{qq}}^{(1)} 
\nonumber \\
&
                \hspace{-.7cm}   
                 -  \beta _0  \bigg( P_{{qq}}^{(1)} +  P_{{qg}}^{(1)}   \frac{ P_{{qq}}^{(0)}}{ P_{{qg}}^{(0)}}  \bigg)      
                + \beta _0  \bigg[   
                -2  C_{2,g}^{(1)} P_{{gq}}^{(0)}       
\nonumber \\
            &
  \hspace{-.7cm}
 +      C_{{2q}}^{(1)}  \bigg( P_{{qq}}^{(0)}     +   P_{{gg}}^{(0)} \bigg)                         
+   C_{2,g}^{(1)} \bigg(   P_{{gg}}^{(0)}     
                   -  P_{{qq}}^{(0)} 
                                \bigg)   \frac{ P_{{qq}}^{(0)}}{ P_{{qg}}^{(0)}}              %
                \bigg]
\nonumber \\
            &
  \hspace{-.7cm}
   + \beta_0^2 \bigg[
             C_{2,q}^{(1)}
              -  C_{\text{2g}}^{(1)}  \frac{ P_{{qq}}^{(0)}}{ P_{{qg}}^{(0)}}   
              \bigg]
%\nonumber \\
%&
 %          
%\nonumber \\
%&
  - \frac{\beta _1}{ \beta _0} \bigg[P_{{gq}}^{(0)} P_{{qg}}^{(0)}   - P_{{gg}}^{(0)} P_{{qq}}^{(0)} \bigg]
\bigg\} 
            \,,
\nonumber \\
K_{DD}^{(1)} & =
              P_{{qq}}^{(1)}  + P_{{gg}}^{(1)} -  \beta_0 \frac{  1 }{P_{{qg}}^{(0)}} P_{{qg}}^{(1)}
+ \beta_1
\nonumber \\
             &
                -2 \beta_0 C_{{2,q}}^{(1)} 
                +   \beta_0 \frac{ C_{{2,g}}^{(1)}}{P_{{qg}}^{(0)}} \bigg(  
                        P_{{qq}}^{(0)}-  P_{{gg}}^{(0)} + \beta_0 
                    \bigg) \; ,
\end{align}
at NLO accuracy to be used in Eq.~(\ref{eq:physsing}). It is immediately obvious from Eq.~(\ref{eq:physsing})
that $K_{22}=0$ to all orders. Likewise, $K_{2D}$ is also trivial and can be read off 
from $dF_2/d\ln Q^2 = K_{2D}\, dF_2/dt$ order by order in $a_s$. 
After properly accounting for differences in the 
terms proportional to $\beta(a_s)$, we agree with the results given in 
\cite{Blumlein:2000wh} where the evolution equation (\ref{eq:physsing})
is expressed in terms of $d/dt$ rather than $d/d\ln Q^2$.
The kernels $K_{AB}$ in (\ref{eq:klo-deriv}) and (\ref{eq:knlo-deriv})
exhibit more moderate higher order corrections, mainly through terms
proportional to $\beta_{0,1}$, than those listed in Sec.~\ref{sec:exp1}.
This shall become apparent in the next Section when we discuss results for
the scale dependence of both $\{F_2, dF_2/dt\}$ and $\{F_2,F_L\}$.

%
%%%%%%%%%%%%%%%%%%%%%%%%%%%%%%%%%%%%%%%%%%%%%%%%%%%%%%
\section{Numerical Studies \label{sec:pheno}}
%%%%%%%%%%%%%%%%%%%%%%%%%%%%%%%%%%%%%%%%%%%%%%%%%%%%%%
%
In this Section we apply the methodology based on physical anomalous dimensions as outlined above
and compare with the results obtained in the conventional framework of scale-dependent
quark and gluon densities and coefficient functions.
Due to the lack of precise enough data for $F_L$ or $dF_2/dt$
we will adopt the following realistic ``toy'' initial conditions for the standard DGLAP evolution
of PDFs at a scale $Q_0=\sqrt{2}\,\mathrm{GeV}$
\cite{Vogt:2004ns} 
\begin{align}
xu_v(x,Q_0^2) &= 5.1072\, x^{0.8}\, (1-x)^3, \notag\\
xd_v(x,Q_0^2) &= 3.06432\, x^{0.8}\, (1-x)^4, \notag\\
x\bar{u}(x,Q_0^2) &=(1-x)\, x\, \bar{d}(x,Q_0^2), \notag\\
x\bar{d}(x,Q_0^2) &=0.1939875\, x^{-0.1}\, (1-x)^5, \notag \\
x\bar{s}(x, Q_0^2) &=0.2\,x\, [\bar{u}(x,Q_0^2)+\bar{d}(x,Q_0^2)], \notag \\
xg(x,Q_0^2) &= 1.7\, x^{-0.1}\, (1-x)^5.
\label{eq:pdfinput}
\end{align}
for all our numerical studies. 
The value of the strong coupling $\alpha_s$ at $Q_0$ is taken to be $0.35$.
For our purposes we can ignore the complications due to heavy flavor thresholds and
set $n_f=3$ throughout. 
We use this set of PDFs to compute the flavor singlet parts of 
$F_2$, $F_L$, and $dF_2/dt$ at the input scale $Q_0$ using Eq.~(\ref{eq:matrixproduct}).
For studies of DIS in the small $x$ region, say $x\lesssim 10^{-3}$, in which we are mainly interested in, the 
flavor singlet parts are expected to dominate over NS contributions
and, hence, shall be a good proxy for the full DIS structure functions. 
Results at scales $Q>Q_0$ are obtained by either solving the RGEs for
PDFs or by evolving the input
structure functions directly adopting Eq.~(\ref{eq:physsing}). For the
solution in terms of PDFs we adopt from now on the standard choice $\mu = Q$.

For completeness and to facilitate the discussions below, let us quickly review the solution of the
matrix-valued RGEs such as Eqs.~(\ref{eq:dglap}) and (\ref{eq:physsing}).
While one can truncate the QCD beta function and the
anomalous dimensions consistently at any given order in $a_s$, there
exists no unique solution to (\ref{eq:physsing})   beyond the LO
accuracy.  The matrix-valued nature of (\ref{eq:physsing}) only allows
for iterations around the LO solution, which at order $a_s^k$ can
differ in various ways by formally higher-order terms of
${\cal{O}}(a_s^{l>k})$.

To this end, we employ the standard truncated solution in
Mellin moment space, which can be found, for instance, in Ref.~\cite{Gluck:1989ze}, see
also \cite{Vogt:2004ns}, and reads
\begin{align}
  \label{eq:NLOtruncated}
  \Gamma_i(n,Q) & =   L_i(n, a_s, a_0)   \Gamma_i(n,Q_0)
\end{align}
where the evolution operator up to NLO is defined as
\begin{align}
 L_i(n, a_s, a_0) & =  L^{(0)}_i(n, a_s, a_0) +  L^{(1)}_i(n, a_s, a_0)
\end{align}
with
\begin{align}
\label{eq:rgesolve0}
 L^{(0)}_i(n, a_s, a_0)  &=  \left( \frac{a_s}{a_0} \right)^{-\frac{\lambda_-}{\beta_0}}
  e_-
 + (+) \leftrightarrow (-)  \notag  \\
%\label{eq:rgesolve1}
 L^{(1)}_i(n, a_s, a_0) &  =  \left( \frac{a_s}{a_0} \right)^{-\frac{\lambda_-}{\beta_0}}
\Bigg[  (a_0 -a_s) e_- R_i^{(1)}  e_-  \notag \\
&  \hspace{-2cm }- \left(a_0 - a_s \left(\frac{a_s}{a_0} \right)^{\frac{\lambda_- - \lambda_+}{\beta_0}} \right)  \beta_0 \frac{e_- R_i^{(1)} e_+}{\lambda_+ - \lambda_- - \beta_0} \Bigg]
\notag \\
\notag \\
&  \qquad
 +  \quad (+) \leftrightarrow (-) \;.
\end{align}
Here, $a_0=a_s(Q_0)$,
$\Gamma_P=\left( \Sigma \atop g \right)$, and
$\Gamma_K= \left( F_A \atop F_B\right)$, i.e., the index $i=P$ refers to the
coupled RGE for the quark singlet and gluon and $i=K$ to the  RGE for the
pair $\{F_A,F_B\}$ of DIS structure functions in (\ref{eq:physsing}).
For $i=K$ one has
\begin{equation}
\left(R_K^{(0)} \right)_{AB} =\frac{1}{\beta_0} K_{AB}^{(0)}\;\;,\;\;
\left( R_K^{(1)} \right)_{AB} =\frac{1}{\beta_0} K_{AB}^{(1)} - \frac{\beta_1}{\beta_0^2} K_{AB}^{(0)}\,,
\end{equation}
with a corresponding definition for $i=P$ in terms of the 
$2\times 2$ matrices of  singlet splitting functions $P^{(0)}$ and $P^{(1)}$.
$\lambda_{\pm}$ denote the eigenvalues given in Eq.~(\ref{eq:eigenvalue}) and
$e_{\pm}$ the projection operators onto the
corresponding eigenspaces; see Refs.~\cite{Gluck:1989ze,Vogt:2004ns}.

As has been mentioned already at the end of Sec.~\ref{sec:exp1}, the eigenvalues
$\lambda_{\pm}(n)$ are identical when computed for the kernels $K_{AB}$ and
$P_{kl}$. This in turn implies that as long as, say, $F_2$ and $F_L$ are calculated at
$\mu=Q_0$ with LO accuracy, their scale evolution based on physical anomalous dimensions 
reproduces exactly the conventional results obtained with the help of
scale-dependent PDFs.

%%%%%%%%%%%%%%%%%%%%%%%%%%%%%%%%%%%%%%%
% FIGURE 3: F2/FL COMPARISON
%%%%%%%%%%%%%%%%%%%%%%%%%%%%%%%%%%%%%%%
%
\begin{figure*}[th]
\begin{center}
\vspace*{-0.6cm}
\epsfig{figure=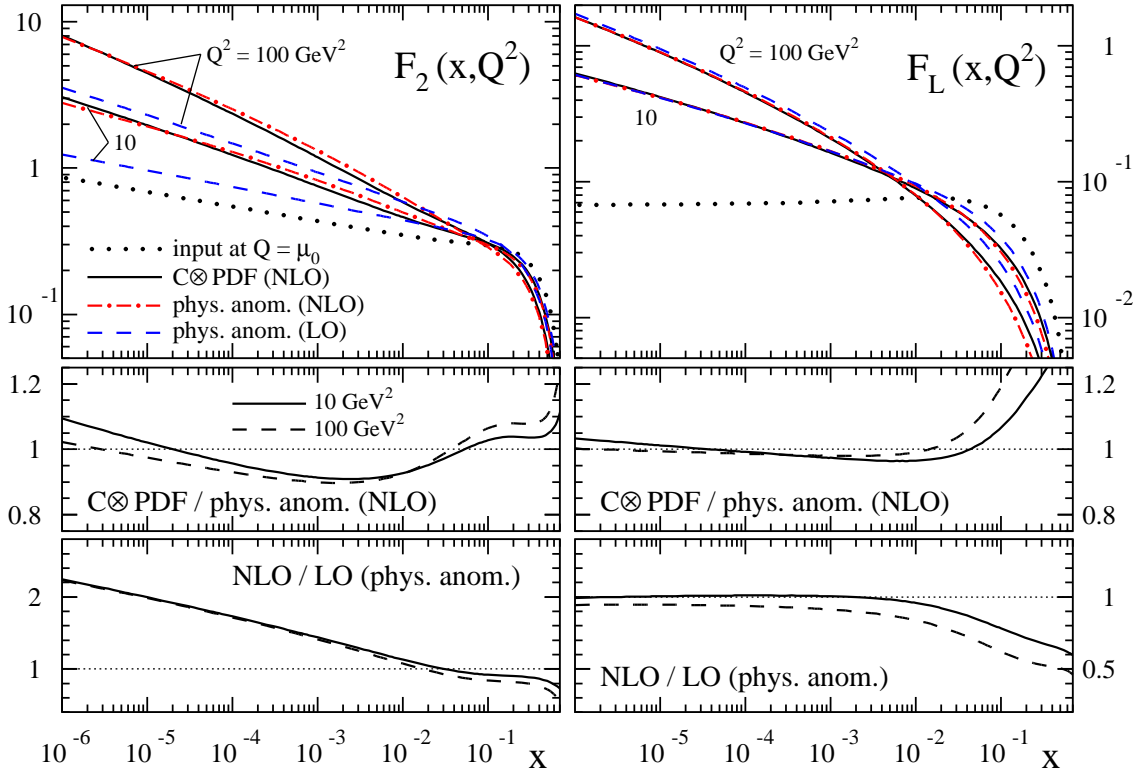,width=0.90\textwidth}
\end{center}
\vspace*{-0.5cm}
\caption{\label{fig:f2fl} [color online] Scale dependence of the DIS structure functions 
$F_2$ {(left)} and $F_L$ {(right)} at NLO accuracy obtained with
physical anomalous dimensions (dash-dotted lines) and in the standard way through a convolution 
of PDFs and Wilson coefficients (solid lines).
The dashed and dotted lines show the results obtained at LO accurary and the input at $Q=Q_0=\sqrt{2}\,\mathrm{GeV}$, 
respectively.  The middle panels give the
ratios of the two different methods to evolve $F_{2,L}$ at NLO, and the lower panels 
illustrate the size of NLO corrections when physical anomalous dimensions are being used;
see text.}
\end{figure*}
Figure~\ref{fig:f2fl} shows our results for the scale dependence of
the DIS structure functions $F_2$ and $F_L$. The input functions at
$Q_0=\sqrt{2}\,\mathrm{GeV}$ are shown as dotted lines.  
While LO results are identical, starting from NLO accuracy the comparison
between the two methods of scale evolution
becomes more subtle, and results {\em seemingly} differ significantly as can be 
inferred from the middle panels of Fig.~\ref{fig:f2fl}.

The origin of the differences between $F_{2,L}(x,Q^2)$ 
computed based on Wilson coefficients and scale-dependent PDFs and 
physical anomalous dimensions can be readily understood from
terms which are formally beyond NLO accuracy.
For instance, upon inserting the NLO
Wilson coefficients (\ref{eq:cexpansion})
and the truncated NLO solution (\ref{eq:NLOtruncated})-(\ref{eq:rgesolve0})
into Eq.~(\ref{eq:matrixproduct}), $F_2$ at ${\cal{O}}(a_s)$
contains spurious terms of both ${\cal{O}}(a_0 a_s)$ and
${\cal{O}}(a_s^2)$.
Since $F_L$ starts one order higher in $a_s$,
similar terms are less important here.
On the other hand, when we
evolve $F_{2,L}$ with the help of physical anomalous dimensions we
first compute, due to the lack of data, the input at $a_0$ based on
Eq.~(\ref{eq:matrixproduct}), which then enters the RGE solution
(\ref{eq:NLOtruncated})-(\ref{eq:rgesolve0}).  
Again, this leads to terms beyond NLO. In case
of $F_2$ they are now of the order ${\cal{O}}(a_0 a_s)$ and
${\cal{O}}(a_0^2)$, i.e., even more relevant than in case of PDFs
since $a_0>a_s$.

To test if the entire difference between the two evolution methods shown in Fig.~\ref{fig:f2fl}
is caused by these spurious higher order contributions, one can easily remove all  
${\cal{O}}(a_s^2)$, ${\cal{O}}(a_0 a_s)$, and ${\cal{O}}(a_0^2)$ contributions from our results.
Indeed, the scale evolution based on physical anomalous dimensions and
the calculation of $F_{2,L}$ from PDFs then yields exactly the same results also at NLO accuracy.
We note that this way of computing properly truncated physical observables from scale-dependent PDFs
beyond the LO accuracy has been put forward some time ago in Refs.~\cite{Gluck:1983bh,Gluck:1991jc}
but was not pursued any further in practical calculations.

Another interesting aspect to notice from Fig.~\ref{fig:f2fl} are the
sizable NLO corrections 
illustrated in the lower panels, in particular, for
$F_2$ in the small $x$ region. 
For this comparison, LO results refer to the same input structure functions
$F_{2,L}$ as used to obtain the NLO results but now evolved at LO accuracy, i.e., 
by truncating the evolution operator in Eqs.~(\ref{eq:NLOtruncated})-(\ref{eq:rgesolve0})
at $L_K^{(0)}$.
At first sight the large corrections appear to
be surprising given that global PDF fits in general lead to acceptable
fits of DIS data even at LO accuracy \cite{ref:pdffits}.
However, this
is usually achieved by exploiting the freedom of having different sets
of PDFs at LO and, say, NLO accuracy to absorb QCD corrections.  
The framework based on physical
anomalous dimensions does not provide this option as the input for the
scale evolution is, in principle, fully determined by experimental
data, and only the value for the strong coupling can be adjusted at
any given order.  In this sense it provides a much more stringent test
of the underlying framework and perhaps a better sensitivity to, for
instance, the possible presence of non-linear effects to the scale
evolution in the kinematic regime dominated by small $x$ gluons.

%%%%%%%%%%%%%%%%%%%%%%%%%%%%%%%%%%%%%%%
% FIGURE 4: F2/F2D COMPARISON
%%%%%%%%%%%%%%%%%%%%%%%%%%%%%%%%%%%%%%%
%
\begin{figure*}[hbt]
\begin{center}
\vspace*{-0.6cm}
\epsfig{figure=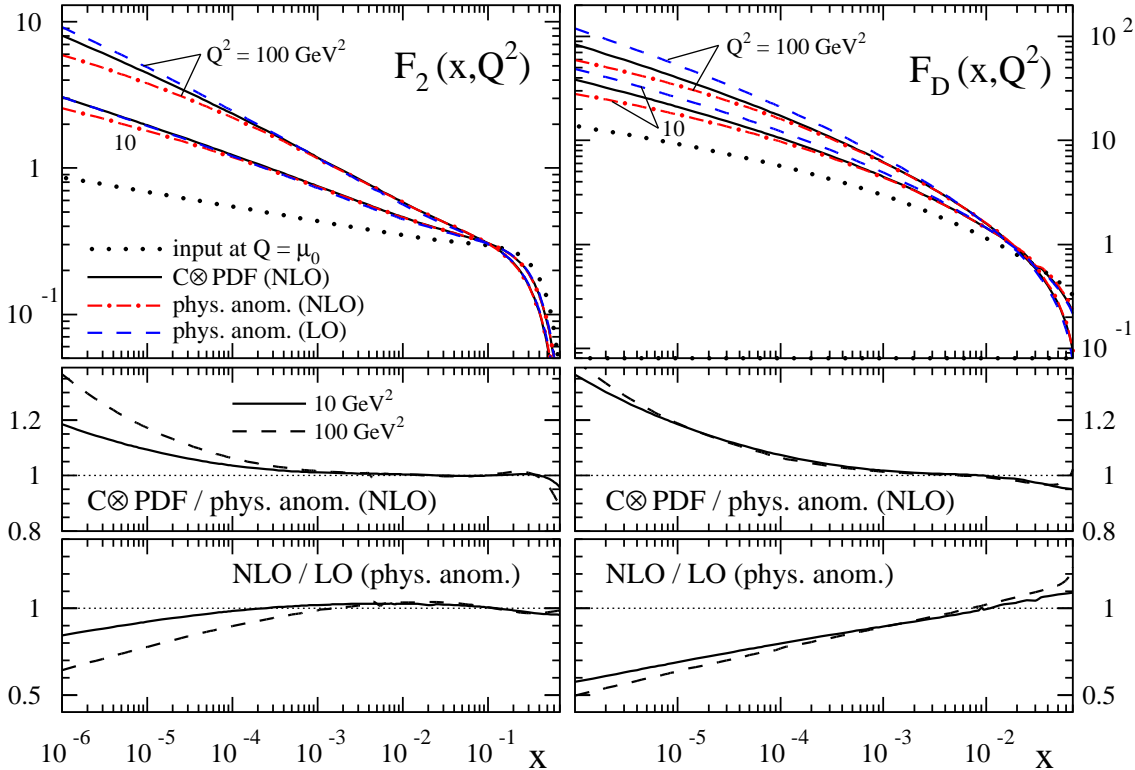,width=0.90\textwidth}
\end{center}
\vspace*{-0.5cm}
\caption{\label{fig:f2fd} [color online] Same as in Fig.~\ref{fig:f2fd} but now for the
pair of observables $F_2$ and $F_D\equiv dF_2/dt$.}
\end{figure*}
In Fig.~\ref{fig:f2fd} we show the corresponding results 
for the scale dependence of the DIS structure function $F_2$ and 
its slope $F_D=dF_2/dt$.
Again, any differences between the scale evolution performed with physical anomalous dimensions
and based on PDFs are caused by formally higher order terms 
${\cal{O}}(a_s^2)$, ${\cal{O}}(a_0 a_s)$, and ${\cal{O}}(a_0^2)$, which can
be removed with the same recipe as above.
As for $\{F_2,F_L\}$, NLO corrections are sizable in the small $x$ region
due to numerically large contributions to $K^{(1)}_{D2}$ and
$K^{(1)}_{DD}$ from the QCD beta function. 

We close our discussions by noting that NLO corrections to the
NS RGEs are much more moderate than in the singlet sector.
This also holds for the importance of spurious higher order 
terms discussed above. Hence, we
refrain from showing any numerical results here.
We also wish to recall that we have ignored for our purposes
any complications in the scale evolution due heavy quark flavors  
by setting $n_f=3$ throughout.
While heavy quarks of mass $m_h$ can be straightforwardly included 
above a scale $Q=m_h$ in a massless approximation
by introducing a new NS combination, which is properly matched
to the singlet part at $Q=m_h$, more sophisticated treatments
require work beyond the scope of this work.
For instance, if heavy quark contributions to DIS structure
functions are treated with their full mass and threshold
dependence, and without resumming any potentially large logarithms
$\log (Q/m_h)$ \cite{Gluck:1993dpa}, one can compute another set of physical anomalous
dimensions for, say, the pair of observables $\{F_2^h,F_L^h\}$,
based on the appropriate fully massive NLO Wilson coefficients in Ref.~\cite{Laenen:1992zk}.

%%%%%%%%%%%%%%%%%%%%%%%%%%%%%%%%%%%%%%%%%%%%%
\section{Summary}
%%%%%%%%%%%%%%%%%%%%%%%%%%%%%%%%%%%%%%%%%%%%%
%
We have presented a phenomenological study of 
the QCD scale evolution of deep-inelastic structure functions 
within the framework of physical anomalous dimensions.
The method is free of ambiguities from choosing a specific
factorization scheme and scale as it does not require the 
introduction of parton distribution functions.
Explicit results for the physical evolution kernels 
needed to evolve the structure functions $F_2$, its $Q^2$ slope,
and $F_L$ have been presented up to
next-to-leading order accuracy.

It was shown that any differences with results
obtained in the conventional framework of scale-dependent
quark and gluon densities can be attributed to the 
truncation of the perturbative series at a given order in the strong coupling.
At next-to-leading order accuracy the numerical
impact of these formally higher order terms 
is far from being negligible but, if desired, such contributions can be 
systematically removed.

A particular strength of performing the QCD scale evolution based on physical
anomalous dimensions rather than using auxiliary quantities such as parton densities
is that the required initial conditions are completely fixed by data and
cannot be tuned freely in each order of perturbation theory.
Apart from a possible adjustment of the strong coupling, this leads to easily testable
predictions for the scale dependence of structure functions and also clearly exposes the relevance of 
higher order QCD corrections in describing deep-inelastic scattering data.
Next-to-leading order corrections have been demonstrated to be numerically sizable,
which is not too surprising given that the physical evolution kernels
absorb all known large higher order QCD corrections to the 
hard scattering Wilson coefficients. 

Once high precision deep-inelastic scattering data from future electron-ion colliders
become available, an interesting application of our results will be to unambiguously quantify the 
size and relevance of non-linear saturation effects caused by an abundance of gluons with small momentum fractions.
To this end, one needs to observe deviations from the scale evolution governed by the physical
anomalous dimensions discussed in this work.
The method of physical anomalous dimensions can be also used for a theoretically
clean extraction of the strong coupling and is readily generalized to other processes
such polarized deep-inelastic scattering or inclusive one-hadron production. 

%%%%%%%%%%%%%%%%
\section*{Acknowledgments}
%%%%%%%%%%%%%%%%
%
We acknowledges support by the U.S.\ Department of Energy under contract number DE-AC02-98CH10886
and by a ``Laboratory Research and Development'' grant (LDRD 12-034) from Brookhaven National Laboratory.

%%%%%%%%%%%%%%%%%%%%%%%%%%%

\end{document}